\documentclass[twocolumn,showpacs,amssymb,aps]{revtex4}

\usepackage{graphicx}
\usepackage{psfig}
\usepackage{dcolumn}
\usepackage{bm}

\def\lessim{\lower.5ex\hbox{$\; \buildrel < \over \sim \;$}}
\def\gtrsim{\lower.5ex\hbox{$\; \buildrel > \over \sim \;$}}
\begin{document} \hbadness=10000
\topmargin -0.8cm\oddsidemargin = -0.7cm\evensidemargin = -0.7cm
\preprint{}

\title{Soft Hadron Ratios at LHC}
\author{Johann Rafelski}
\affiliation{Department of Physics, University of Arizona, Tucson, Arizona, 85721, USA}
\author{Jean Letessier}
\affiliation{Laboratoire de Physique Th\'eorique et Hautes Energies\\
Universit\'e Paris 7, 2 place Jussieu, F--75251 Cedex 05
}
 
\date{May 30, 2005}

\begin{abstract}
High precision  soft hadron abundance data
produced in  relativistic nuclear collisions at LHC  
at $\sqrt{s_{\rm NN}}\le 5500$ GeV will become available
  beginning in 2007/8. We explore, within the statistical hadronization
model, how these results can help us understand  the  
properties of the deconfined quark--gluon phase at its breakup.  
We make assumptions about the physical properties
of the fireball and obtain particle production predictions. 
Then, we develop a strategy
to measure parameters of interest, such as strangeness 
occupancy $\gamma_s$,  chemical potentials $\mu_{\rm B}$ 
and $\mu_{\rm S}$. 
\end{abstract}

\pacs{24.10.Pa,  25.75.-q}
\maketitle
\section{Introductory Remarks}
\label{intro}
Relativistic  heavy ion collisions experimental program has as 
objective the formation of the deconfined quark--gluon plasma 
(QGP) phase in the laboratory.   
 The  uncertainty in this experimental program is if, in the available  short
collision time, $10^{-22}$--$10^{-23}$~s, the color frozen nuclear phase 
can melt and turn into the QGP state of matter. There is
no valid first principles answer available today, nor it seems,
it can be expected   in the foreseeable future. From 
this, and other such uncertainties about the QGP, arises the need 
to define and  study its  observables, even though we are 
quite convinced that this is the state of matter that
filled the Universe in its early stage, till hadronization
occurred  10--20$\mu$s into  its evolution. QGP is the 
 {\it equilibrium} state in a hot Universe 
at   temperature above that of lightest hadronic particle, the pion,
$kT>m_\pi c^2=140$ MeV (we hence use units such that $k=c=\hbar=1$). 

Detailed theoretical study of the properties of this new  state of matter
shows that QGP is rich in entropy and strangeness. These
are the observables discussed here explicitly, and implicitly, 
in the context of soft hadron production.
The enhancement of entropy $S$ arises in the early stages of 
the collision process,
because the color bonds are broken, and numerous gluons are formed
and  thermalized. Enhancement of
 strangeness $s$ is in part also due to the breaking of 
color bonds. Furthermore, it is  due to  a modification of the
 kinetic strangeness formation processes. These
operate  faster in deconfined phase, mainly because  the 
mass threshold for strangeness excitation is considerably lower 
in QGP than in hadron matter, but also because there are 
more channels available considering the color quantum numbers.

We will not discuss the following important QGP observables in this work,
their current status at RHIC   is discussed, {\it e.g.\/},
in Ref.\,\cite{Huang:2005nd}.   Aside from strangeness and entropy 
enhancement, another soft hadron  signature is  
the shape of particle spectra, which  carries  information
about the formation and evolution dynamics of the state 
of matter that is the source  of these particles. 
Given the considerable increased energy, we expect a greater 
energy density in the initial stage, and thus a   much more 
violent transverse outflow of matter than has been seen at RHIC. 
Such a strange transverse collective flow   carries
many particles to high transverse momenta, and produces  a strong 
azimuthal asymmetry in particle spectra for finite impact parameter
reactions. Among other 
related hadronic signatures, we note a significant charm
quark abundance, originating in primary parton  reactions.
The pattern of charm hadronization should reveal further 
details about the QGP phase, just as strange hadrons do. 
 
Within a few years, a  new energy domain will become accessible 
in study of heavy ion collisions, when
the Large Hadron Collider (LHC) at CERN becomes operational in 2007.
The top energy available to Pb--Pb reactions is $\sqrt{s_{\rm NN}}=5500$ GeV,
a 27.5 fold increase compared to the top RHIC energy. Extrapolating the trends
of SPS and RHIC physics, we expect a much greater entropy and strangeness
yields at central rapidity. We will always address, in
this work, most central  head on collision  reactions of two $A\simeq 200$
heavy nuclei, at LHC this will be Pb--Pb collision . 

At these high energies,  there 
will be much less stopping power of baryon number, and thus the 
central rapidity region will be much more similar to the phase prevailing
in the early Universe than this is the case at RHIC. One of the objectives
of this work is to assess how small the baryochemical potential $\mu_{\rm B}$
can become, compared to $\mu_{\rm B}\simeq  25 $ \,MeV observed at RHIC. 
The scale of $\mu_{\rm B}$ at the hadronization of the early Universe
is  $\mu_{\rm B}^{\rm U}\simeq 1$ eV~\cite{Fromerth:2002wb}.

In this paper, we address specifically the pattern of soft hadron production
based on the assumption of a sudden breakup of the deconfined hadron 
phase with all soft hadrons produced at essentially same 
physical conditions, and  not subject to requirement of the 
absolute chemical equilibrium condition. We will, however,  provide
reference data for the chemical equilibrium   case.
In the next section \ref{SHM}, we outline the statistical hadronization
model and present its parameters. In section \ref{constrain}, we discuss 
how simple, but general, hypothesis allow to fix values of these parameters.
We establish range  of physical interest for strangeness
phase space occupancy $\gamma_s$. 
In section \ref{predict},
we develop our predictions regarding  the properties of
the hadronizing fireball. We present the range of statistical
parameters which we expect and the physical properties of the 
fireball at its breakup. We consider  particle ratios  which could
help determine the value of  $\gamma_s$, which is a parameter 
in this study. 
In section \ref{mubsec},
we  consider observables sensitive to  $\mu_{\rm B}$, and obtain 
results which show how precise the hadron yields need to be
measured in order to allow measurement of $\mu_{\rm B}$.

\section{Statistical Hadronization and Model Parameters}
\label{SHM}  
Statistical Hadronization Model (SHM) is, by definition, a model  of 
particle production in which the formation  process of each particle 
fully saturates (maximizes) the quantum mechanical probability amplitude. 
Particle yields are thus determined by the appropriate integrals of the
accessible phase space \cite{JJBook}.
 For a system subject to global dynamical 
evolution, such as collective flow,  this    applies
within each volume element, in its local rest  frame of reference. The  SHM  
is  consistent with the wealth of SPS and  RHIC data available today.
Systematic study of particle production for 
a wide reaction energy range confirms applicability of 
the SHM,   see  \cite{Letessier:2005qe,BDMRHIC}.

Analysis of hadron yields further   facilitates a study  of the 
physical properties of the hadronic fireball   at the  time 
of hadronization, when these particles 
are produced, {\it i.e.\/}, undergo chemical freeze-out. 
A  study of hadron multiplicities, 
produced at energy   ranging from the top of AGS energy to the top
of RHIC energy, leads to an understanding of the 
physical properties of the fireball at hadronization~\cite{Letessier:2005qe}.
Here, we reverse the approach --- using  
the systematics of the  energy dependence of the physical properties of the
fireball, we establish our
expectations about the   statistical parameters and 
 relative  particle multiplicities 
expected at LHC.  We cannot address, in this work, the 
total hadron yield, since this depends on the early stage of 
the reaction, and specifically, on the entropy formation process
 in initial  interactions. 
 
The complexity of the SHM model derives from the need to 
account for many hadronic resonances, and their decays.
Since the number of massive resonances grows exponentially,
their contributions to particle yields, especially to the yields
of pions, are slowly convergent. Moreover, the counting of 
resonances and their conserved
 quantum numbers poses a significant book keeping 
challenge. For this reason, an effort has been made to generate
a comprehensive software package available to all interested
parties, with a transparent hadron data input, and a comprehensive
parameter field. All results   presented
were obtained using this numerical package SHARE 
(Statistical Hadronization with REsonances)~\cite{share}.

Another package of similar capability, but with restrained 
parameter set, has since become available~\cite{Wheaton:2004qb}.
These programs are including a large number of resonances
and track the chemical composition as well as 
the decay trees  with care. As result,  
the benchmark fits produce a hadronization temperature which 
is considerably lower than obtained in Ref.\,\cite{BDMRHIC} for
SPS or RHIC reaction systems.  

A successful description of  rapidity particle yields 
within the SHM, at a  single-chemical
freeze-out condition, produces   the model 
parameters in the process of $\chi^2$ minimization. The
parameters are:\\
1) $dV/dy$,  the volume  
related a given rapidity to the particle yields;\\
2) $T$, the  (chemical) freeze-out temperature;\\
3) $ \mu_B\equiv T\ln (\lambda_u \lambda_d)^{3/2}$,  the baryon    and \\
4) $\mu_S\equiv T\ln [\lambda_q/\lambda_s]$,  hyperon  chemical potentials;\\
5) $\lambda_{I3}\equiv\lambda_u/\lambda_d$, a fugacity distinguishing 
the up from the down quark flavor;\\
6) $\gamma_s$ the strangeness phase space occupancy;\\
7) $\gamma_q$ the light quark phase space occupancy.\\
When the assumption of absolute chemical equilibrium is made the values 
$\gamma_s=\gamma_q= 1$ are set. The relationship to quark flavor fugacities  
$\lambda_i, i=u,d,s$ is made explicit above. 

When a set of parameters is known, all particle multiplicities can be 
evaluated exactly. Similarly,  one can obtain the physical properties
of the fireball such as thermal energy, entropy, baryon content by
appropriate evaluation of the properties of partial fraction contributions
of each hadronic state. In that way, in fact, one obtains from a fit to a
limited set of measured
particle yields a full  phase space extrapolation for any particle
yield, and a full understanding
of the properties of the fireball. Some striking `conservation of fireball
properties' rules emerged from such an analysis of particle yield
data~\cite{Letessier:2005qe},  and these we will use in order 
to predict the (relative) particle yields at LHC energy range. 

\section{LHC Hadronization}
\label{constrain}
 \subsection{Choice of conditions and  constraints}
\label{statpar}
To predict the 7 parameters, at first sight, we need at least 7, and
better more than that,  valid conditions, constraints and hypothesis:\\
1) The `volume' normalization $dV/dy$ only enters absolute
hadron yields. Consequently, it is related to the initial conditions,
{\it i.e.\/}, mechanisms of entropy production. Restricting our investigation 
to the  study of  particle ratios, we do not need
to know the value of this parameter, which normalizes the overall yield.  

There are two natural  physics constraints:\\
2) Strangeness conservation, {\it i.e.\/}, the (grand canonical)
 count of $s$ quarks in  hadrons equals   $\bar s$ count at 
each rapidity unit. In our specific case, we request that: 
\begin{equation}
{\bar s-s\over \bar s+s}=0\pm0.01.
\end{equation}
We will show  how this condition
establishes the relationship between the baryochemical
potential $\mu_{\rm B}$ and the strangeness chemical 
potential  $\mu_{\rm S}$ in some detail in subsection \ref{scon}.\\
3) The electrical charge to net baryon ratio, 
in the final state, should be the same as in the initial state, and 
in the specific case of Pb--Pb interactions, we have: 
\begin{equation}
{Q\over b}=0.39\pm 0.01.
\end{equation}
The 2.5\% error can be seen as expressing uncertainty about how
well neutron and proton densities follow each other, given that even
most central reactions occur at finite impact parameter, and 10\%
of nucleons do not participate in the reaction.\\
4) and 5)  

i) The    phase space occupancy parameters   in some approaches 
are tacitly fixed:  assuming  the chemical equilibrium   one sets
$$\gamma_s=\gamma_q=1\,.$$  

ii) In the chemical non-equilibrium approach, the
systematics of   data analysis at RHIC and high SPS energies 
firmly predicts:
$$\gamma_q\simeq e^{m_{\pi^{\!{\rm o}}}/2T}.$$
 Furthermore, we 
will present our results as function of $\gamma_s$. We believe
that the value of $\gamma_s$ is linked to the collision energy, as 
more strangeness can be produced, when the initial conditions reached 
become more extreme. We  show, in section \ref{results},
how $\gamma_s$ fixes several easily accessible
observables and can be measured, and the consistency of the 
chemical non-equilibrium approach checked. At LHC energy, the expected
value of $\gamma_s$ is so much greater than unity, and thus, we can 
be sure that a distinction from  $\gamma_s=1$  for the equilibrium model
  can be arrived at in an unambiguous way. 

There are (at least)  two further   conditions required which must be sensitive
 to the   hadronization temperature and baryochemical potential.
These will be drawn from the following  observations:\\
{6)} We note the value, 
$${E\over TS}\to 0.78,\quad {\rm or}\to 0.845, $$ 
for  chemical non-equilibrium~\cite{Letessier:2005qe},
or respectively, equilibrium model analysis.
We note that the energy per particle
 of  non-relativistic and semi-relativistic classical 
particle gas is $E/N\simeq m+3/2\, T+\ldots$, while
the entropy per particle 
in this condition is $S/N=5/2+m/T+\ldots$ (see section 10 of \cite{JJBook}).
Hence:
\begin{equation}\label{ETS}
{E\over TS}\simeq {m/T+3/2\over m/T+5/2}.
\end{equation}
It is thus possible to interpret
this constraint  in terms of a quark matter made of particles with 
thermal mass  $m\propto aT$. Solving Eq.\,\ref{ETS}, we find for
${E/TS}= 0.78$, $a=2$ for chemical non-equilibrium,
and for
${E/TS}=0.845$, $a=4$ for equilibrium. This is near to the result expected in 
finite temperature QCD~\cite{Petreczky:2001yp}.
 That result points to a simple structure of the   quark matter
fireball. On the quark-side, the value $E/TS$ 
is   not very model specific, though
it is  sensitive  to the average particle mass as shown above. 

On the other hand,  there is considerable sensitivity to 
this thermodynamic constraint
in the hadronic gas. The hadron system
after hadronization comprises   a mix of particles of different, and
for the baryon component,  large mass. To fine tune this value a 
specific ratio of baryons to mesons needs to be established: in this
way the hadron system can maintain both energy and entropy, aside of 
baryon number and strangeness, during hadronization. For this reason, in 
the hadron phase there 
is considerable sensitivity of this ratio to both   $T$ and the 
value of the phase space occupancy parameters, here  $\gamma_s$.  \\
{7)}  We   need to  make an assumption which fixes the baryochemical
potential $\mu_{\rm B}$. This certainly is one of most difficult guesses as
there is no reliable way to predict baryon stopping at LHC, and certainly this
value, 
$$\mu_{\rm B}\ll T,$$ 
will be quite difficult to measure. For this reason,
we will discuss, in subsection \ref{mubsec}
at length a method to measure  $\mu_{\rm B}$.  
 The impact of $\mu_{\rm B}\simeq 1$--3  MeV
  on mixed particle ratios, such as ${\rm K}/\pi$ or $\eta/\pi^0$ is 
  physically irrelevant. We will show haw particle--antiparticle 
abundances can help us further. In fact, it is uncertain that
a measurement of such small $\mu_{\rm B}$ can be accomplished,  and 
thus in effect, we could have simply assumed  $\mu_{\rm B}=0$ which at 
a few \%-level 
would be consistent with all relative hadron
yield predictions here presented. On the other hand, the 
understanding of matter--antimatter asymmetry present at LHC
energy scale is, in itself, of interest and thus, we pursue this 
question further. 

  We argue  as follows: our   analysis
of RHIC data suggested that baryons are more easily retained in the 
central rapidity region than energy, with 
2.5 times larger fraction of colliding baryons  than fraction of energy deposited.  
 Considering the SPS data point, and the RHIC results, 
the  per baryon thermalized reaction energy 
 retained in the central rapidity region, 
 in units of the maximum available collision energy, drops from 40\% 
available at RHIC 
 to   15\% at LHC.   Given the   LHC energy  $\sqrt{s_{\rm NN}}=5500$ 
GeV,  a thermal energy  content per net baryon  at $y=0$ is assumed to be,
 $${dE\over db}=0.15 \times 5500/2=412\pm 20 \,{\rm  GeV} ,$$
 at top LHC energy, where the error is chosen to be similar 
in relative magnitude as the error in other observables considered.
This error plays a role in finding the solution in terms of statistical
variables of the constraints and conditions considered.
 
This specific assumption fixes implicitly the (small) value of the 
baryochemical potential, and by virtue of the strangeness conservation,
also of the strange chemical potential. When we deviate from this
assumption in exploring a wider parameter range, we will mention 
this explicitly.

\subsection{$\eta$ and maximum value of $\gamma_s$}
\label{limitgammas}

We will study the hadronization condition as 
function of $\gamma_s$ which may take large values. We note that 
$\gamma_s$ cannot rise above a  limit, to be
determined   from   similar consideration
as is the maximum value of 
\begin{equation}
\gamma_q^{\rm CR}=\exp(m_\pi/2T).
\end{equation} 
At this value, the Bose distributions of pions diverges.  
As   $\gamma_s$  increases, same will happen in the strange hadron sector
and  indeed this will occur first to 
the lightest particle with considerable hidden
strangeness content, {\it i.e.\/}, $\eta(548)$. Naively, 
one could expect that $\gamma_s^{\rm CR}=\exp(m_\eta/2T)$. However, 
$\eta$, unlike the spin 1 $\phi(1020)$ is not  a 
(nearly) pure $\bar s s $ state. 

The quark structure  of 
$\eta(548)$ and $\eta'(958)$  can be written as:
\begin{eqnarray}
\eta&=&{u\bar u+d\bar d\over  \sqrt{2}}\cos \phi_p  - s\bar s \sin\phi_p,\\
\eta'&=&{u\bar u+d\bar d\over  \sqrt{2}}\sin\phi_p  + s\bar s \sin\phi_p.  \nonumber
\end{eqnarray}
Study of numerous experimental results, and in particular of  $Z^0\to $ hadrons, 
shows that   $ \sin^2\!\!\phi_p=0.45\pm 6$ indicating that $\eta(548)$
has 45\%  strangess content~\cite{Uvarov:2001wv}. 
This arises from the 
SU(3)-flavor octet state content of 67\% reduced by SU(3)-symmetry
breaking mixing with the 33\% strangeness content of the singlet $\eta'(958)$. 

In order to count the yield 
of the $\eta$, we introduce its fugacity $\Upsilon_\eta$. The 
fractional contribution to the partition function is:
\begin{equation}
\ln Z_\eta=- \int dV \!\!{d^3p\over (2\pi)^3} 
         \ln\left(1-\Upsilon_\eta e^{-E_\eta\over T}\right),
\end{equation}
with $E_\eta=\sqrt{m_\eta^2+p^2} $. We focus our attention  
on the dominant, directly produced $\eta$. The incremental 
per unit volume $\eta$ yield is:
 \begin{equation}
  {dN_\eta\over dV}
= \Upsilon_\eta {\partial[d\ln Z_\eta/dV] \over\partial\Upsilon_\eta}
= \int\!\! {d^3p\over (2\pi)^3} 
 {1\over{ \Upsilon_\eta^{-1} e^{E_\eta\over T}-1}}.
\end{equation}
More specifically, the rapidity density is:
 \begin{equation}
n_\eta\equiv {dN_\eta\over dy} = {dV\over dy}\times {dN_\eta\over dV}.
\end{equation}

The `normalization' $dV/dy$ which comprises the transverse dimension
at hadronization, and the longitudinal incremental volume, is 
arising from kinetic expansion processes driven by the initial state formation
mechanisms. These have to be obtained by methods beyond the scope of this work. 
However, we note that  in the longitudinally scaling limit for ideal fluid 
 hydrodynamic evolution of the initial state,
 at all times  the entropy rapidity density remains constant, 
$dS/dy={\rm Const.}$. Since the particle multiplicity is defined by the 
value of $dS/dy$, the total
hadron multiplicity is not affected by our ensuing   
  study of hadronization for different chemical freeze-out temperatures.
The total charge particle rapidity density  is  a consequence of initial 
processes which are beyond the physics reach of this work.
 
We now relate  the $\eta$-fugacity $\Upsilon_\eta$ to the light and strange
quark fugacities $\gamma_q$ and $\gamma_s$. 
In the SHM, the  probability of the production  of 
 $\eta$ is weighted  with the  yield of strange
 and light quark pairs in proportion of their 
contribution to the quark content in the particle formed.
Thus,
\begin{equation}\label{etafug}
\Upsilon_\eta = \gamma_q^2 \cos^2 \phi_p + \gamma_s^2 \sin^2 \phi_p <e^{m_\eta/T}.
\end{equation}
The upper limit is set at the phase space divergence point. Specifically, 
considering $\eta(548)$ with its 45\% strangeness content, we find from  Eq.\,(\ref{etafug}) a maximal value $\gamma_s<10.4$ for the hadronization
conditions of interest in this work, {\it i.e.}, $\gamma_q\to e^{m_\pi/2T} $ and 
$T\to 140$ MeV. Thus we set as the range of interest $0<\gamma_s<10$.  

This upper limit is the most stringent constraint for  $\gamma_s$:
it is more stringent than the one which 
arises from  the consideration of the $\phi$-meson, 
\begin{equation}\label{phifug}
\Upsilon_\phi = \gamma_s^2  <e^{m_\phi/T},
\end{equation}
due to the greater $\phi$-mass. Similarly, the  constraint on  $\gamma_s$ 
 based on the kaon condensation (in presence of negligible chemical potentials):
\begin{equation}\label{Kfug}
\Upsilon_{\rm K} = \gamma_s\gamma_q   <e^{m_{\rm K}/T}.
\end{equation}
is less severe. 
 
To confirm the functional dependence on $\gamma_i$, Eq.\,(\ref{etafug}),   
we show that it
 is consistent with the requirement that the fugacity $\gamma_i$,
allows the count of valance quark content in hadrons. 
We look at $\gamma_s$, which
allows the count of all strange and antistrange quarks by the relation:
\begin{equation}
s +\bar s = \gamma_s {\partial \ln Z\over \partial \gamma_s}.
\end{equation}
The contribution of $\eta$ to the strangeness count thus is, 
\begin{eqnarray}
(s+\bar s)_\eta &=& {\gamma_s\over\Upsilon_\eta} 
    {\partial\Upsilon_\eta\over \partial \gamma_s}
 \Upsilon_\eta {\partial\ln Z_\eta \over\partial\Upsilon_\eta},
\nonumber\\
&=&2 {\gamma_s^2 \sin^2 \phi_p\over 
         \gamma_q^2 \cos^2 \phi_p + \gamma_s^2 \sin^2 \phi_p} N_\eta,
\end{eqnarray}
which is the required result. A similar procedure can be followed to 
show that:
\begin{equation}
(q+\bar q)_\eta = 2 {\gamma_q^2 \cos^2 \phi_p\over 
         \gamma_q^2 \cos^2 \phi_p + \gamma_s^2 \sin^2 \phi_p} N_\eta.
\end{equation}
We   conclude that
 Eq.\,(\ref{etafug}) defines the 
$\eta$-fugacity in terms of light and strange quark fugacities. 

\subsection{Range of $\gamma_s$ of interest}
When $\gamma_s$ increases, the  $\eta$-fugacity increases rapidly, 
but, even at $\gamma_s=5$, it is well below the Bose singularity. For
$T=160$ MeV, the singularity would be just above $\gamma_s=8$. However,
the fall-off of the expected hadronization temperature (see below)
moves this towards twice as large value. One  may 
wonder how large a  value of $\gamma_s$ is physically 
consistent, and in particular, if an  associate   decrease in the value 
of $T$   makes sense.

In qualitative  terms, this type of parameter evolution and correlation  is
fully consistent with the picture of rapid transverse expansion of the QGP 
fireball. This expansion can lead to supercooling which  pushes the 
hadronization temperature lower. At the same time, the preserved yield of 
strangeness requires that $\gamma_s$ increases. 

We now look at the possible range of expected values 
of $\gamma_s$ in more detail. It can be expected   that, for LHC extreme
conditions, the   strangeness  phase space has been chemically 
saturated   at a temperature {\em larger} than strange quark mass: $T_1>m_s$.
Moreover, in the   QGP phase, there is a residual strange quark
mass $m_s>T$, where $T$ is the final hadronization temperature. While
the precise value of $m_s$ depends on the momentum scale at which it is measured,
we will assume in this semi-quantitative discussion that
 $m_s(1 \mbox{ GeV})\simeq 180$ MeV.  Conservation
of strangeness rapidity yield in the expansion from $T_1$ to $T$ implies,
by comparison of the relativistic phase space size,
\begin{equation}\label{gamsQ}
(dV_1T_1^3 )W(m_s/T_1)=\gamma_s^{\rm Q}\times (dVT^3)W(m_s/T) ,
\end{equation}
where $\gamma_s^{\rm Q}$ is the quark-side hadronization phase space occupancy,
for definition of $W(x)$, see Eq.\,(\ref{defF}). For $0<x<1$, $W(x)$ is slowly 
changing and close to its asymptotic value $W(x=0)=2$. Entropy conservation 
further requires that $  dV_1T_1^3=dVT^3=\mbox{Const.}$

For $T_1\simeq 1.5 T$, and noting that $T_1>m_s> T$,  
we  obtain,  as solution of Eq.\,(\ref{gamsQ}),
 $\gamma_s^{\rm Q}\simeq 1.5$.   
In the hadronization process, this value increases by a significant factor which 
we can obtain comparing the phase space of strange hadrons in QGP with that
hadron phases (we omit as is customary the upper index H, {\it i.e.\/},
$ \gamma_s^{\rm H}\to \gamma_s$):
\begin{eqnarray}
\label{gamsH}
\gamma_s^{\rm Q}\times W(m_s/T)&=&\gamma_s \times\sum_i\tilde W(m_i/T)\\ \nonumber
   &+&\gamma_s^2 \times \sum_j\tilde W(m_j/T) \\ \nonumber
   &+&\gamma_s^3 \times \sum_k\tilde W(m_k/T). 
\end{eqnarray}
The sums run over single, double and triply strange ($s$ and $\bar s$)
hadrons, respectively. 
$\tilde W$ differs from $W$ in that it comprises appropriate hadron
 fugacities for each particle,  the appropriate expressions 
are seen in more detail
in Eq.\,(\ref{lnZs}). 

The  strangeness QGP to HG 
aspect ratio,   {\it i.e.\/}, the ratio of the phase phase space
size seen in  Eq.\,(\ref{gamsH}) up  to the coefficients  
$\gamma_s^{\rm Q}$, $\gamma_s$  
shows that it  grows with decreasing temperature, see Fig.\,19.3 in 
Ref.\,\cite{JJBook}. This is   further strongly amplifying by a factor
as large as four  the final observed 
$\gamma_s$. A QGP phase 
value  $\gamma_s^{\rm Q}\simeq 1.5$  may become
 $\gamma_s \simeq 5$--10  on hadron side.
 As we shall see, the decrease of $T$ with 
increasing $\gamma_s$ compensates, in part, the increase with $\gamma_s$  
in  the yields of  strange hadrons.   We  will thus consider the range
$0.5<\gamma_s<10$ as that is where most of variation in the considered
observables is seen, and the actual physical conditions are expected
to occur. The lower limit underscores the comparison with the 
equilibrium model $\gamma_s=\gamma_q=1$ behavior. The upper limit $\gamma_s<10$
is within the range of allowed values of $\gamma_s$ we have obtained in 
previouse section \ref{limitgammas}.

\section{Predictions}\label{predict}
\subsection{SHM parameter values at LHC top energy}
With the assumptions outlined in subsection \ref{statpar}, 
we solve for the best set of SHM model parameters.
A precise solution is always found in the considered range of
 $\gamma_s$, also 
for the increased value of $E/TS$ for the case of chemical equilibrium. 
This of course does not guarantee that, {\it e.g.\/}, our 
baryochemical potential is correctly chosen. The results we
show are not entirely smooth as we allow a small error in the
constraints and conditions, and thus the solution for the parameters
is not a precise algebraic result, but a most likely value of 
a quasi-fit, which has a $\chi^2\simeq 0$.

\begin{figure}[!bt]
\hspace*{-.6cm}
\psfig{width=8.5cm,figure=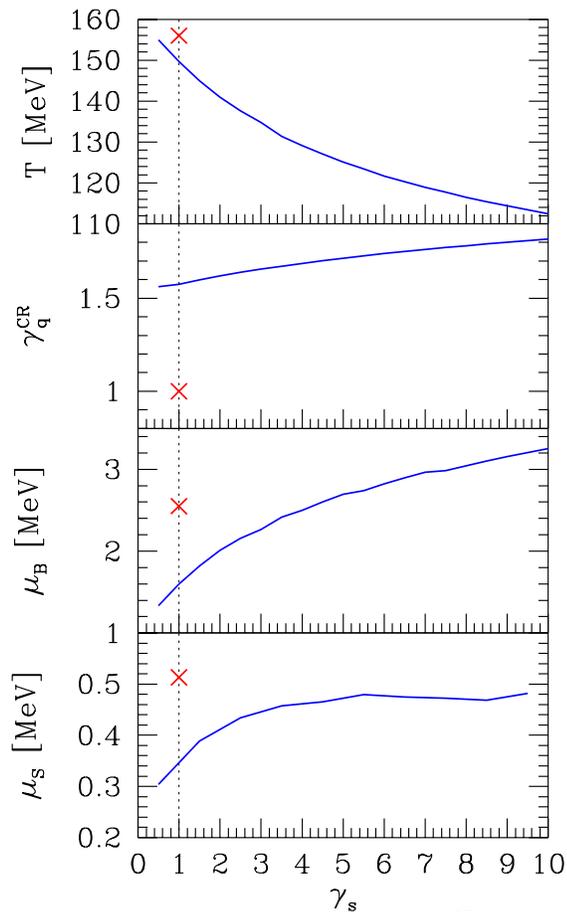}
\vskip -1.cm
\caption{\label{STATPAR}
(color online) The values of $T$, $\gamma_q^{\rm CR}$, 
$\mu_{\rm B}$, and $\mu_{\rm S}$ 
as function of varying $\gamma_s$, consistent with the hadronization 
model assumptions outlined in subsection \ref{statpar}. The equilibrium 
model results are crosses   at $\gamma_s=1$ for  $\gamma_q=1$.
}
\end{figure}

The resulting statistical parameters $T$, $\gamma_q^{\rm CR}$, 
 $\mu_{\rm B}$, and $\mu_{\rm S}$ 
 are shown, in Fig.\,\ref{STATPAR},
as function of  $0.5\le \gamma_s\le 10$.  Note that through our study of 
chemical non-equilibrium
  $\gamma_q\simeq\gamma_q^{\rm CR}$.
We see that an   increasing value of $\gamma_s$
is accompanied by   considerable reduction of the  
hadronization temperature, which drops from the value   $T=140$ MeV,
near $\gamma_s=2.4$ and $\gamma_q=1.6$,  down to $T=110$~MeV. At
the favorite  value $\gamma_s=5$, the expected LHC hadronization temperature
is  $T=125$ MeV for the  chemical non-equilibrium. The chemical 
equilibrium result at $T=156$ implies a chemical potential 
$\mu_{\rm B}=2.6$ MeV, and this value is also found      for
the chemical non-equilibrium model for $\gamma_s=5$. We explain why
$\mu_{\rm B}$ remains unchanged at the end of section \ref{mubsec}.
For $\mu_{\rm S}$, we see a slight 
reduction by 10\% from equilibrium model value at 0.52 MeV as is appropriate
considering the results presented in Fig.\,\ref{muSmuB}. Our estimate of 
the chemical parameters at LHC are considerably different 
from those proposed by others~\cite{Andronic:2003zv}, 
where $\mu_{\rm  B}=1$ MeV
and $\mu_{\rm  S}=0.3$ MeV is proposed. We note that  our 
baryochemical potential is considerably greater. We will discuss 
experimental  consequences
at greater length in  section \ref{mubsec}.

\begin{figure}[!bt]
\hspace*{-.6cm}
\psfig{width=8.5cm,figure=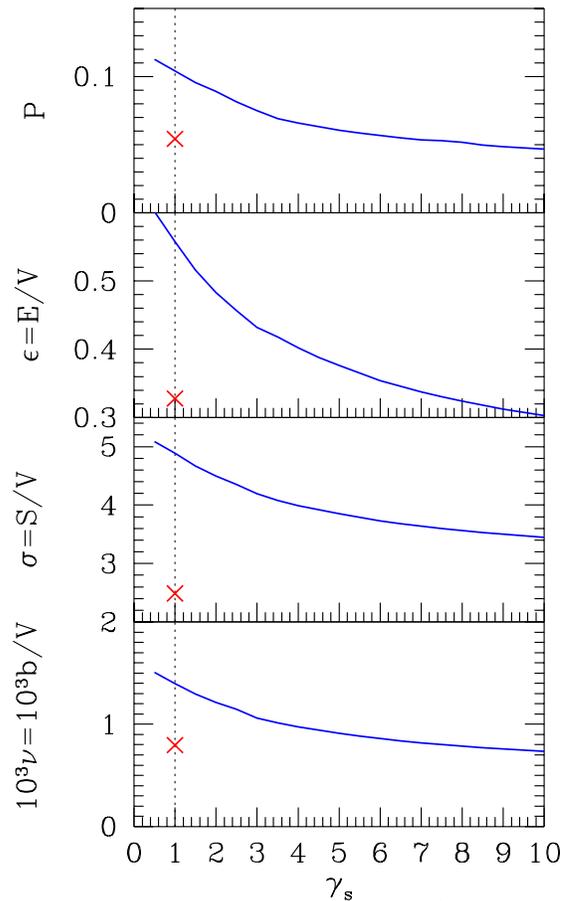}
\vskip -1cm
\caption{\label{PHYS}
(color online) Pressure $P$ [GeV/fm$^3$], 
energy density $\epsilon$ [GeV/fm$^3$], 
entropy density $\sigma=S/V$ [1/fm$^3$], 
net baryon density $\nu= (B-\overline B)/V=b/V $ [1/fm$^3$], 
for non-equilibrium SHM. Cross at $\gamma_s$ for 
chemical equilibrium. 
}
\end{figure}

It is of considerable interest to study the   physical properties
at hadronization:  the pressure $P$, the energy density $\epsilon=E/V$, 
entropy density $\sigma=S/V$, and net baryon density $\nu=B/V$. 
In general, 
the non-equilibrium hadronization occurs from a state of higher density,
as is seen  in Fig.\,\ref{PHYS}, comparing the lines with the SHM equilibrium 
cross. This is, in particular, true for the entropy density. 
 With increasing  $\gamma_s$,  all 
density decreases, the drop in temperature is a more important 
influence than the large increase in relative strangeness   yield. 
Despite a significant increase in $\mu_{\rm B}$ with increasing $\gamma_s$, the 
net baryon  density decreases modestly. It is very small, bordering
 the value $\nu=0.001$ fm$^{-3}$.

\subsection{Particle yield ratios and determination of $\gamma_s$}
\label{results}

\begin{figure}[!bt]
\hspace*{-.6cm}
\psfig{width=8.5cm,figure=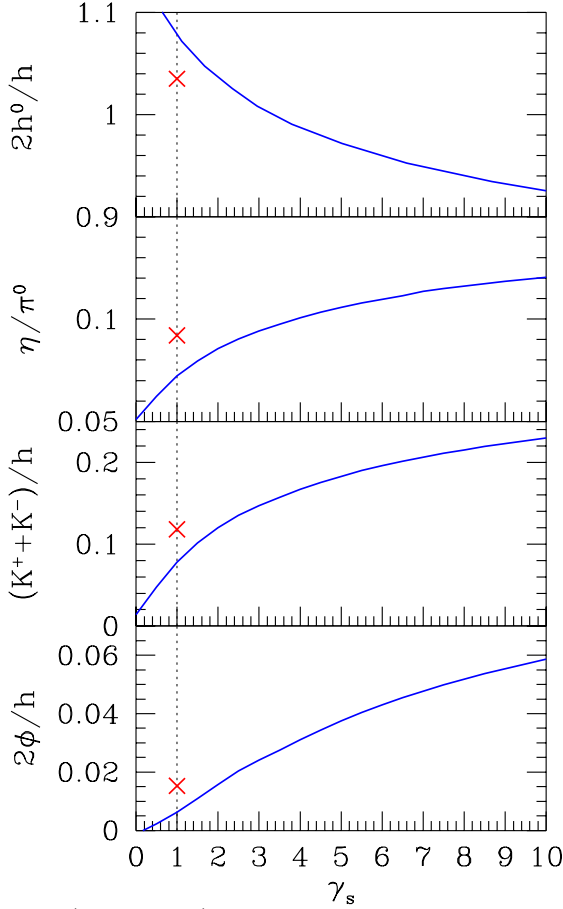}
\vskip -1.cm
\caption{\label{partrat}
(color online) The predicted yield ratios as function 
of $\gamma_s$, from top to bottom: ratio of neutral to charged hadrons
$2h^0/h$, 
 $\eta/\pi^0$, $({\rm K}^++{\rm K}^-)/h$ and $2\phi/h$ 
as function of $\gamma_s$. The cross indicates chemical equilibrium model
prediction. All yields after weak decay of hyperons and ${\rm K}_{\rm S,L}$.
 }
\end{figure}

 There is approximate charge symmetry with positives $h^+$ and 
negatives $h^-$ having a very similar yield. The difference
will be discussed further below. The total charged 
hadron yield will be denoted as, 
$$h=h^++h^-\equiv p+\bar p+\pi^++\pi^-+{\rm K}^++{\rm K}^-,$$
and is evaluated after weak decay of hyperons and ${\rm K}_{\rm S,L}$.
Similarly the total yield of neutrals:
$$h^0\equiv \pi^0+n+\bar n.$$ 
The top panel, in Fig.\,\ref{partrat}, shows the $2h^0/h$ ratio, which varies
by $\pm10$\% in the range of $\gamma_s$ considered, with the charge symmetric
 value $2h^0/h=1 $ arising at $\gamma_s=3.3$.

In the lower three panels, we focus our 
interest on ratios of some interest for determination of $\gamma_s$. 
We present the  ratio $\eta/\pi^0$ which rises  by nearly
 50\% compared to the equilibrium 
model expectation, see crosses at $\gamma_s=1$, computed for 
$\gamma_q=1$ and $T=156$ MeV. 
This ratio is observed by reconstruction of the invariant di-photon mass.
This observable shows a relatively small variation with $\gamma_s$, 
which can be understood as result of competition between $\gamma_s$ and $\gamma_q$, see
Eq.\,(\ref{etafug}), which is accompanied by the
 decrease of $T$ with increasing $\gamma_s$. We record 
that the expectation for the non-equilibrium hadronization at LHC is: 
$$0.07< {\eta\over \pi^0}<0.12\,.$$ 
This interesting observable may also not have the sensitivity required
to distinguish the models, or help determine $\gamma_s$.

In the panel below, we show  the  $({\rm K}^++{\rm K}^-)/h$ 
ratio.  This ratio is rising, for  
 large $\gamma_s$,  to a value near 0.23  almost double 
 the `standard' chemical equilibrium value at 0.12.
 However, for $\gamma_s=5$, we see a more modest increase  by 50\%. We
record for LHC:
$$ 0.12<  {{\rm K}^++{\rm K}^-\over h^++h^-}< 0.23 \,.$$
More spectacular is the expected increases in the $2\phi/h$ ratio.  The
Chemical equilibrium value of 0.015 is seen to rise 4-fold, and 
at $\gamma_s=5$, we still see a very noticeable increase  by 
a factor 2.5.  This is a very important observable of 
the condition of hadronization. We record our LHC expectation:
$$ 0.015 < {2\phi\over  h^++h^-}<0.06 \,.$$
  
\begin{figure}[!bt]
\hspace*{-.6cm}
\psfig{width=8.5cm,figure=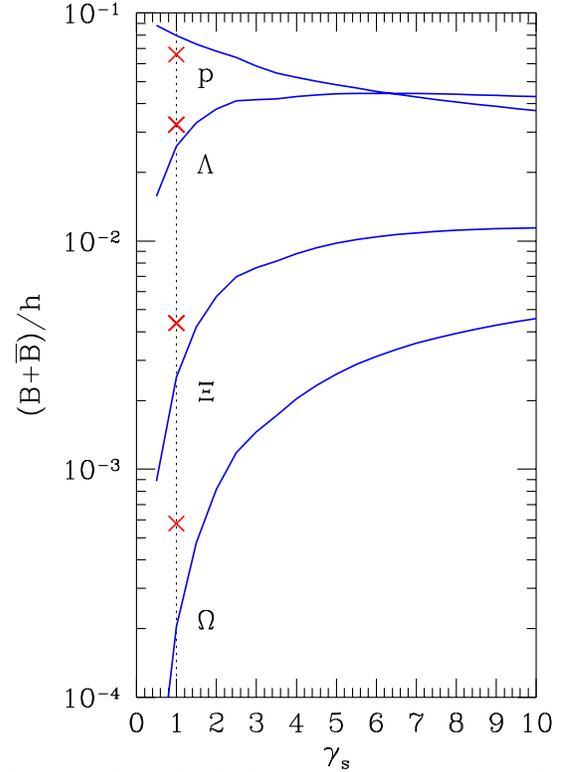}
\vskip -1.cm
\caption{\label{baryrat}
(color online) (Strange) baryon ratios with charge 
hadron multiplicity $(p+\bar p)/h$,\ $\Lambda+\overline\Lambda)/h$,\ 
$(\Xi^-+\overline\Xi^+)/h$ and $(\Omega^-+\overline\Omega^+)/h$, 
after all weak decays occurred. Crosses denote chemical equilibrium
result. }
\end{figure}

The behavior of the baryon yields is shown in Fig.\,\ref{baryrat} where the 
total, nearly matter--antimatter 
symmetric yields of protons, and the three hyperon families are shown, normalized
by the total charged hadron yield $h$. The lines, from top to bottom
are for $(p+\bar p)/h$, $(\Lambda+\overline\Lambda)/h$,
$(\Xi^-+\overline\Xi^+)/h$ and $(\Omega^-+\overline\Omega^+)/h$. 
We note that for large $\gamma_s$, a
considerable change in the expected baryon populations ensues, with 
proton yield decreasing (due to decrease in $T$ while the more strange
the hyperon is, the more it is enhanced compared to 
chemical equilibrium expectations.  Interestingly, these relative
 yields saturate with increasing $\gamma_s$, as the effect of temperature
decrease competes with the increase due to rising $\gamma_s$. 
Note that we have   shown,
in Fig.\,\ref{baryrat},
the total hadron yields after all weak decays have occurred. We summarize our LHC 
expectations:
\begin{eqnarray}
0.07>&{\displaystyle p+\bar p\over\displaystyle h^++h^-}&>0.04,\nonumber\\
0.02<&{\displaystyle\Lambda+\overline\Lambda\over\displaystyle h^++h^-}&<0.04,\nonumber\\
0.004<&{\displaystyle\Xi^-+\overline\Xi^+\over\displaystyle h^++h^-}&<0.015,\nonumber\\
0.0006<&{\displaystyle\Omega^-+\overline\Omega^+\over\displaystyle h^++h^-}&<0.004.\nonumber
\end{eqnarray}

\section{Measurement  of the baryo-chemical potential}
\label{mubsec}
 \subsection{Strangeness conservation}
\label{scon}
An interesting challenge, at LHC, will be the measurement of 
chemical potentials. We recall 
that $\mu_{\rm S}$ is directly related to $\mu_{\rm B}$
should  hadron emission at each rapidity occurs such that 
there is local strangeness balance, {\it i.e.\/}, `conservation'. 
This relates  the two chemical potentials as we shall next discuss.
Otherwise, if emission of hadrons were to reflect on
the QGP conditions, we would expect $\mu_{\rm S}=0$ and 
this would generate a buildup of residual strangeness in a 
distillation process~\cite{destil}. We tacitly assumed, and continue
in this way now, that at LHC,  in each 
rapidity region, local conservation of strangeness prevails. 
We present a set of results which will
allow, given appropriate experimental sensitivity, to determine the 
value of $\mu_{\rm S}$.  

Strangeness conservation establishes a relation between the chemical 
potentials~\cite{ChemCons}.
While this relationship is simplified for the  case
 $\mu_i/T\ll 1$, the presence of 
$\gamma_s\gg 1$ introduces new elements and we reinspect the 
relationship.  Using, at first, the quark fugacity notation
for convenience, the open strangeness
sector partition function is, in the 
here appropriate Boltzmann approximation:
\begin{eqnarray}
\ln Z_s &\equiv& \gamma_s\gamma_q F_{\rm K} 
  \left( {\lambda_s\over\lambda_q} +{\lambda_q\over \lambda_s}\right )
   + \gamma_s\gamma_q^2 F_{\rm Y}  
      \left(\lambda_s \lambda_q^2+{1\over \lambda_s\lambda_q^2}\right)
 \nonumber\\  \label{lnZs}
&&\hspace*{-0.7cm} + \gamma_s^2\gamma_q  F_{\Xi}
      \left(\lambda_s^2 \lambda_q +{1\over \lambda_s^2\lambda_q }\right)
  + \gamma_s^3  F_{\Omega}
      \left(\lambda_s^3 +{1\over \lambda_s^3 }\right).\quad
\end{eqnarray}
The phase space factors, 
\begin{equation}\label{defF}
\hspace*{-0.2cm}F_i(T)={VT^3\over 2\pi^2}\sum_{k\in i} g_k W(m_k/T), \  W(x)=x^2K_2(x),
\end{equation}
 comprise all contributing hadron states `$k$' with  
quantum number `$i$'.   $W(x)=x^2 K_2(x)\stackrel{x\to 0}{\to} 2$ is  the
relativistic phase space integral in classical limit, which for large $x$ behaves
 as $W(x)\propto x^{3/2}\exp(x/T)$. A series of these terms represents quantum
statistics, see Eq.\,(\ref{murel}).

\begin{figure}[!bt]
\hspace*{-.6cm}
\psfig{width=9.5cm,figure=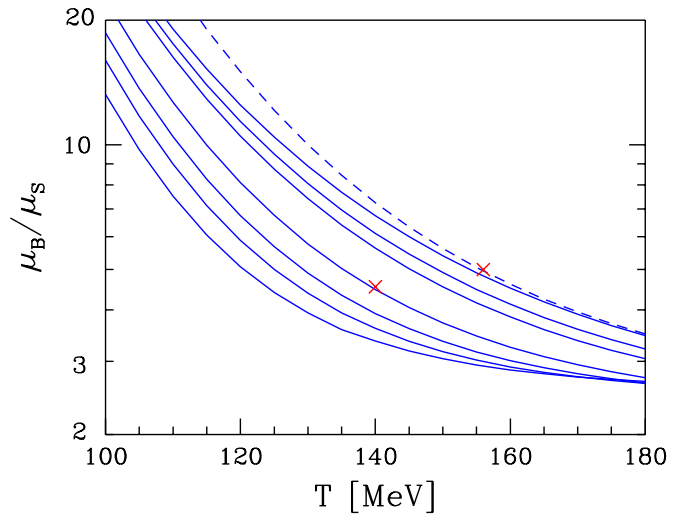}
\vskip -0.4cm
\caption{\label{muSmuB}
(color online)  $\mu_{\rm B}/ \mu_{\rm S}$  
  as function of $T$: From top right to left
  $\gamma_s=0,\ 0.5,\  1,\ 3,\ 5,\ 7,\  10$, at $\gamma_q=\exp({m_{\pi^0}/T})$.
The dashed (red) line  shows the chemical  equilibrium 
model result    at $\gamma_s=1$ and  $\gamma_q=1$. Crosses correspond
to RHIC freeze-out conditions both in equilibrium (at dashed line)
and non-equilibrium. 
}
\vskip -0.3cm
\end{figure}

The  strangeness conservation condition,
$$\langle s\rangle-\langle \bar s\rangle
  ={\lambda_s\partial \ln Z\over \partial \lambda_s}=0,$$ 
yields for small values of chemical potentials,  in units of~$T$:
\begin{eqnarray}
\label{murel}
&&\hspace*{-0.7cm}\mu_{\rm B}\!-\mu_{\rm S}
  +2(\mu_{\rm B}\!-2\mu_{\rm S}){\gamma_s\over \gamma_q}{F_\Xi \over  F_{\rm Y}}
 + 3 (\mu_{\rm B}\!-3\mu_{\rm S}){\gamma_s^2\over \gamma_q^2}{F_\Omega \over  F_{\rm Y}} \  
\\ \nonumber
&=& \mu_{\rm S}\left( {F_{\rm K}\over \gamma_qF_{\rm Y}}
  +{  \gamma_ s\tilde F_{\rm K}(2m_{\rm K})\over 4 F_{\rm Y}}  
  +{ \gamma_q\gamma_s^2\tilde F_{\rm K}(3m_{\rm K})\over 9 F_{\rm Y}}  +\ldots \right )\,.
\end{eqnarray}
 The right hand side presents the three first terms
of the kaon  Bose integral  expansion which keeps terms of the same 
order as those belonging
to $\Xi$ and $\Omega$ in the balance. 
These are not entirely negligible for large $\gamma_s$. 
We solve Eq.\,(\ref{murel}) and show, in
 Fig.\,\ref{muSmuB}, $\mu_{\rm B}/ \mu_{\rm S}$
as function of $T$:
\begin{equation}
 \label{murelval}
{\mu_{\rm B}\over \mu_{\rm S}}=  f(T;\gamma_q,\gamma_s).
\end{equation}
The dashed curve is the equilibrium case for $\gamma_s=\gamma_q=1$, 
where we marked the RHIC hadronization condition $\mu_{\rm B}/ \mu_{\rm S}= 5$
with a cross. The closest solid line to this result is for  
 $\gamma_q=e^{m_{\pi^0}/T}$ and $\gamma_s\to 0$. The 
following solid lines are,
in sequence from upper right, for  $\gamma_s=  0.5,\, 1,\, 3,\, 5,\, 7,\, 10$.
The cross near the $\gamma_s=3$ line,  at
  $\mu_{\rm B}/ \mu_{\rm S}=4.6$,  corresponds to the 
RHIC chemical non-equilibrium hadronization condition.

These results allow, aside of gauging the LHC values of chemical
potentials,  also an easy comparison with and cross check of  other work
addressing LHC and RHIC environments with conserved strangeness. 
We recall that many particle ratios are directly determined
by chemical potentials, {\it e.g.\/}, ${\rm K}^+/{K^-}=\exp(2\mu_{\rm S}/T)$
and $\overline\Xi^+/\Xi^-=\exp((2 \mu_{\rm B}-4\mu_{\rm S})/T)$. This obviously
leads to consistency conditions in the particle--antiparticle asymmetry
sensitive particle ratios.

\subsection{Hadron--antihadron asymmetry}

In order to measure chemical potentials, we need to be 
able to measure the particle--antiparticle asymmetry.
This is not an easy task at LHC as we shall see.
Up to RHIC energy, it was customary to study ratio of 
antiparticle yields to particle yields, such as $\overline\Lambda/\Lambda$.
At LHC, the strategy has to slightly change.
We consider of normalized particle--antiparticle  difference  yields:
\begin{equation}
\Delta N_i ={{\bar n_i-n_i}\over{\bar n_i+n_i}}, 
\end{equation} 
where $n_i$ is the rapidity density $dN/dy$ of charged particles 
$i={\rm K}^{+},\ p,\ \Lambda,\ \Xi,\ \Omega$ and antiparticles. We omitted
intentionally the pion from this list, as it is the dominant component
of the unidentified particle hadron asymmetry, we will address next.
Below, we will omit the factor $d/dy$ as the expressions we state are valid 
more generally.  

\begin{figure}[!bt]
\hspace*{-.6cm}
\psfig{width=9cm,figure=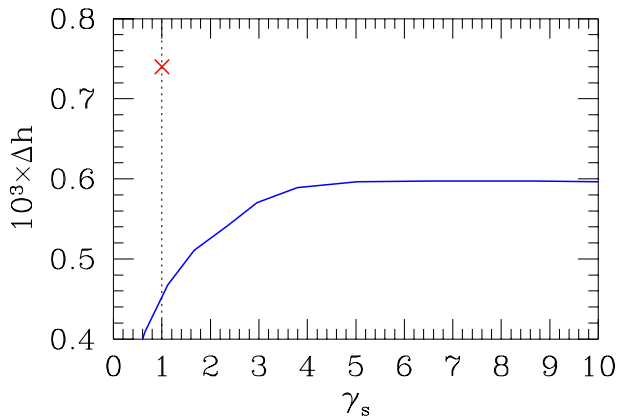}
\vskip -0.6cm
\caption{\label{dh}
$\Delta h$ as function of $\gamma_s$.
Cross indicates equilibrium model result.
}
\end{figure}

The   most accessible observable,
\begin{equation}\label{delh}
\Delta h\equiv {h^+-h^-\over h^++h^-},
\end{equation}
composed of  unidentified charged hadrons  is very
hard to measure precisely, for i) there are 
distortion  possible by partial acceptance of weak decays, 
and ii) this variable assumes a comparatively  small value   at LHC. Moreover,
this is  also an observable hardest to interpret in a simple and transparent 
theoretical model. We  find that its
  magnitude for  LHC will be:
$$\Delta h = (0.5\mbox{--} 0.6 )10^{-3},$$
for all values of $\gamma_s\ge 1$, as is 
shown in Fig.\,\ref{dh}. The difference in 
$\Delta h$  between equilibrium and non-equilibrium result
is  due to the increased entropy and thus hadron multiplicity content 
of the chemical non-equilibrium case. The variable $\Delta h$ has been 
recognized already in the study of SPS reactions as a sensitive probe of 
entropy production~\cite{ChemCons,entro}.

\begin{figure}[!bt]
\hspace*{-.8cm}
 \psfig{width=9.5cm,figure=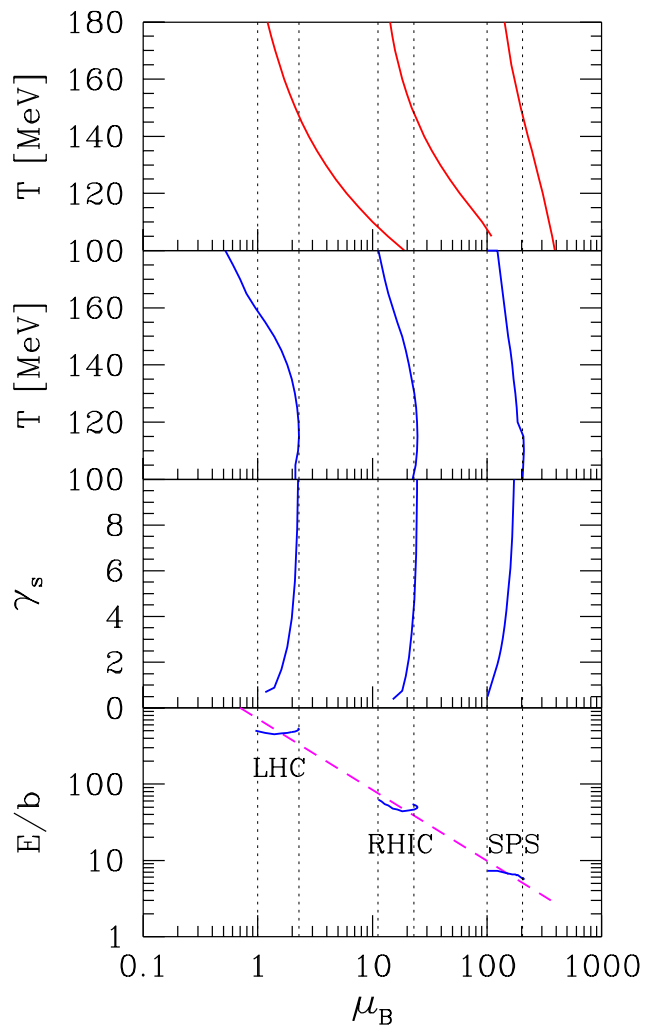}
\vskip -0.5cm
\caption{\label{Dh}
(color online)  $\Delta h$ constraint 
from left to right  for LHC   $\Delta h=0.001$; for RHIC   $=0.01$  
 and for SPS top energy  $=0.1$. 
 Top panel: chemical equilibrium SHM, $T$--$\mu_{\rm B}$ plane, 
bottom three panels: chemical non-equilibrium, from top to bottom  $T$,
followed by $\gamma_s$ and $dE/db$ combined with $\mu_{\rm B}$. In the bottom panel 
the dashed line indicate the systematics of the behavior regarding
the value of thermal energy per baryon. Vertical dotted lines brace the 
extreme allowable values of $\mu_{\rm B}$.
 }
\end{figure}

Should a way to measure $\Delta h $ of this 
magnitude with reasonable    precision 
be found at LHC, this   would provide  a  model
 independent measure of the value of the baryo-chemical potential 
$\mu_{\rm B}$. To see this, we solve (fit) the SHM  with
 a fixed assumed  value of $\Delta h$. In the chemical
equilibrium version of SHM, we find, at $\gamma_s=\gamma_q=1$, in the 
$T$--$\mu_{\rm B}$ plane the constraint lines determined by assumed 
values of $\Delta h$, shown   in the top panel of Fig.\,\ref{Dh}. 
From left to right, we have considered three benchmark values, 
$\Delta h=0.001 $ (top LHC energy),
$\Delta h=0.01 $ (top RHIC energy), and 
$\Delta h=0.1 $ (top SPS energy, central rapidity region). At SPS, 
top energy in S--Pb interactions,  we 
have studied this variable  (called $D_Q\simeq 0.009$) in the context of 
the exploration of 
the specific entropy per baryon~\cite{ChemCons,entro}, and the 
results we present here are in agreement with this early study 
of SPS hadron multiplicities.

For non-equilibrium case, with $\gamma_q= \gamma_q^{\rm CR}$,
  the corresponding  result is shown in 
the second top panel. The nearly vertical lines are solution 
of the the following conditions imposed:
 strangeness conservation, charge-to-baryon ratio and $\Delta h$,
as function of $\mu_{\rm B}$.
For  each value of $T$, there is a specific value of $\gamma_s$ 
indicated in the next lower panel, and again, we see nearly a 
vertical line. Both these results imply that while the value of 
$T$ and $\gamma_s$ are not much constrained  by the measurement of 
just one single observable $\Delta h$, the value
of $\mu_{\rm B}$ is already highly constrained. 

 Why this is the
case is understood inspecting   
the bottom panel in Fig.\,\ref{Dh}. We show the
resulting value of thermal energy per baryon $E/b$. 
These turn out to be  highly
localized regions. Thus, $\Delta h$ is for a wide range of 
other statistical parameters closely related to the value of 
energy per baryon, or equivalently, entropy per baryon. 
 The   connecting dashed line in the bottom
panel guides the eye.   A check of E/TS also confirms that these
solutions produce the expected result which is otherwise introduced as a 
constraint. 

Thus, we learn that just the single `measured'   value  $\Delta h$ is enough to 
constrain  a  rather  narrow range of $\mu_{\rm B}$. This result is indicated
by the vertical dotted line, which we place bracing  the domains of 
$\mu_{\rm B}$ that are allowed. Inspecting the equilibrium model intercept,
we realize that this singles out a domain of $T$  which is result of 
data fits with this constraint. 
This explains   why $\mu_{\rm B}$ is usually determined
in a model independent way within the SHM, with little if any difference
present  between the different  model variants, provided that 
the experimental 
data used in the fit comprises explicitly, or implicitly, $\Delta h$.
For example, the study of the impact parameter dependence at RHIC using 
different SHM model variants produced $\mu_{\rm B}$ and $\mu_{\rm S}$ which
cannot be distinguished (see bottom panel of Fig.\,1 in Ref.\,\cite{bdepend}).

\subsection{Identified particle--antiparticle asymmetries}
For the identified particles, the normalized
 particle--antiparticle   differences can be closely and analytically related
to the  value of chemical potential. For example, the kaon asymmetry 
is   directly related to strangeness chemical potential $\mu_{\rm S}$: 
  \begin{equation}\label{must}
\Delta {\rm K}\equiv {{{\rm K}^+-{\rm K}^-}\over{{\rm K}^+ + {\rm K}^-}}
              \simeq \tanh {\mu_{\rm S}\over T }
              \to  {\mu_{\rm S}  \over T}.
\end{equation}
We have, for simplicity, not considered the $\phi$-meson
 decay contributions which
 increase the normalizing   yield but do not alter the difference. 

\begin{figure}[!bt]
\hspace*{-.6cm}
\psfig{width=9.5cm,figure=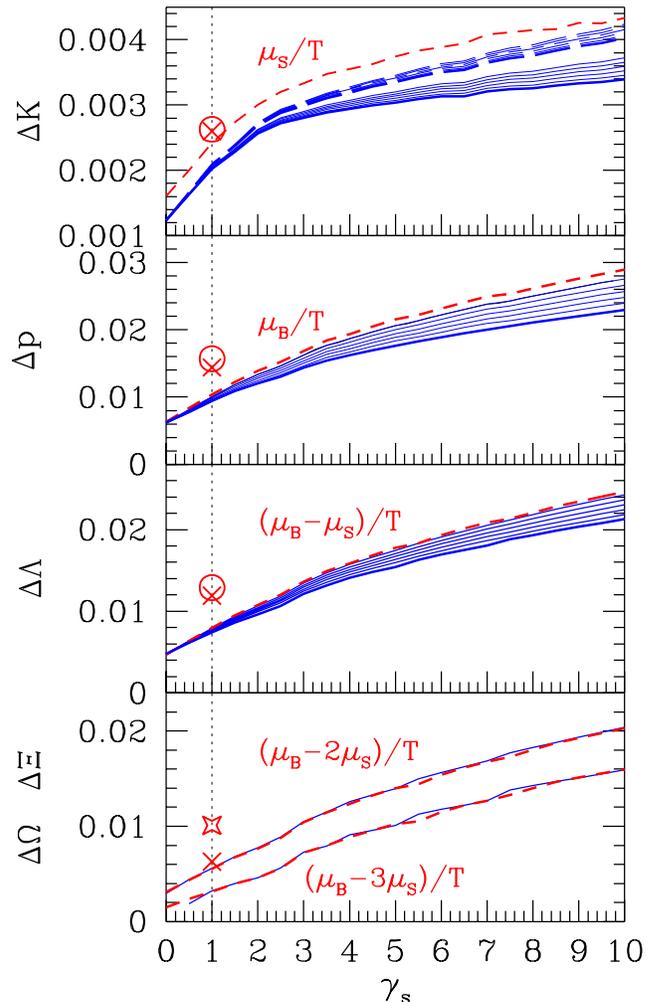}
\vskip -1.cm
\caption{\label{mumeas}
(color online) Solid lines: the relative particle antiparticle asymmetry 
as function of $\gamma_s$, from top to bottom $\Delta {\rm K}$, $\Delta p$, 
$\Delta \Lambda$ and together in bottom panel 
 $\Delta \Xi$ and  $\Delta \Omega$. The short 
dashed line (red) is the analytical result, given in terms of $\mu_i/T$
(see text). The range of weak decay corrections is shown by parallel lines with
the most decay contaminated result being the bottom, thick line. For $\Delta K$
we also show by a long dashed lines the result after the $\phi$-decay contribution
to $\Delta K$ has been removed. The equilibrium model results are shown
as crosses (after weak decays) and circles (corrected for weak decays)
at the $\gamma_s=1$ vertical dotted line. The 
diamond symbol in bottom panel is the chemical equilibrium result for $\Delta \Xi$,
different from the cross for $\Delta \Omega$. 
}
\end{figure}

We show the actual, with all decays, $\Delta {\rm K}$ as thick 
solid line  in the top panel of Fig.\,\ref{mumeas} (bottom line in this top
panel). The very top short dashed line in the panel (red on-line) 
is the ratio $\mu_{\rm S}/T$. 
The thick long-dashed line excludes from the ratio the 
contamination by the decay
$\phi\to {\rm K}^++{\rm K }^-$. The fully weak
decay contamination corrected results are the thin (solid and resp. dashed)
lines at the top of the parallel lines, and are shown for both
the full result (solid lines) and $\phi$-decay corrected result (long dashed).
The parallel line regions are where the acceptance 
of weak decays is partial and/or the correction is incomplete. After the
removal of the $\phi$-decay dilution of the kaon yields,  
Eq.\,(\ref{must}) should read:
  \begin{equation}\label{must1}
\Delta {\rm K} \simeq  0.9 {\mu_{\rm S}  \over T}.
\end{equation}
The  slight reduction from the analytical formula   Eq.\,(\ref{must})  
 is due to the strong decay contributions of hyperon
resonances decaying emitting a kaon. Because of the smallness of $\Delta K$, the
baryon asymmetry in the hyperon resonances leaves this small but visible 
 imprint of this  result.  

The chemical equilibrium result (cross before and circle after weak decays)
is also indicated in   Fig.\,\ref{mumeas}. These are, in general, 
larger than the analytical results (short dashed lines) except in the case of 
$\Delta K$.

For baryons there are four particle--antiparticle 
 differentials, which are shown below $\Delta K$, 
in Fig.\,\ref{mumeas}. We expect for protons:
\begin{equation}\label{mub}
\Delta p\equiv {{p-\bar p}\over{p+\bar p}} =\tanh {\mu_{\rm B}\over T}
              \to {\mu_{\rm B} \over T}.
\end{equation}
The thin solid line in the second panel from the top
which corresponds to removed weak decays 
compares well to the analytical results, 
short-dashed line. The thick solid line at bottom of parallel lines includes in 
$\Delta p$ the contamination from weak decays 
of  $\Lambda$ and $\overline\Lambda$, and the
region in between  spans all possible  WD contamination.  

Once  the weak decay contributions  
$\Omega,\Xi\to \Lambda$ and $\overline\Omega,\overline\Xi\to \overline\Lambda$
are removed, we have further:
\begin{equation}\label{mubl}
\Delta \Lambda\equiv {{\Lambda-\overline\Lambda}\over{\Lambda+\overline\Lambda}}
          = \tanh {\mu_{\rm B}-\mu_{\rm S} \over T}
              \to {{\mu_{\rm B}-\mu_{\rm S}} \over T}.
\end{equation}
The thin line, in the $\Delta \Lambda$ panel,
is nearly indistinguishable from this
result. The thick solid line includes all weak decays.  

There is no significant contamination of $\Xi^-$ and $\overline\Xi^+$ and 
thus we have:
\begin{equation}\label{mubx}
\Delta \Xi\equiv {{\Xi^--\overline\Xi^+}\over{\Xi^-+\overline\Xi^+}}
         =\tanh {\mu_{\rm B}-2\mu_{\rm S} \over T}
              \to {{\mu_{\rm B}-2\mu_{\rm S}} \over T},
\end{equation}
and similarly for the $\Omega^-$:
\begin{equation}\label{mubo}
\Delta \Omega\equiv {{\Omega^--\overline\Omega^+}\over{\Omega^-+\overline\Omega^+}}
         =\tanh {\mu_{\rm B}-3\mu_{\rm S} \over T}
              \to {{\mu_{\rm B}-3\mu_{\rm S}} \over T},
\end{equation}
both shown in bottom panel of Fig.\,\ref{mumeas}.

For baryons, the expected asymmetry is at \%-level and  
the weak decay of hyperons allow unique 
identification of these particles. It is 
quite possible that the measurement of these 
variables will succeed. 
 As the above expressions show, there is a consistency condition
since two chemical potentials and the temperature 
 considering  that the isospin bath of numerous pions causes 
$\lambda_{I3}\to 1$ to a great precision, such 
that $T\ln\lambda_{I3}\ll\mu_{\rm S}$ determine five 
observables $\Delta N$. This consistency is 
further tightened     due to strangeness conservation relation
of $\mu_{\rm B}/\mu_{\rm S}=4$--5, see section \ref{scon}.  
The strangeness conservation constraint and through it
the SHM model   can be tested by transforming
Eqs.\,(\ref{must},\ref{mub},\ref{mubl},\ref{mubx},\ref{mubo}) so we 
can make use of  
Eq.\,(\ref{murel}) and the values Eq.\,(\ref{murelval}):
\begin{eqnarray}
{\Delta p \over   \Delta {\rm K} }
   &=& {\mu_{\rm B}\over \mu_{\rm S}}  \simeq 4.5  ,\\
{\Delta \Lambda \over  \Delta {\rm K} } 
   &=& \left( {\mu_{\rm B}\over \mu_{\rm S}}-1\right)   \simeq 3.5    ,\\
{\Delta \Xi \over  \Delta {\rm K} } 
   &=& \left( {\mu_{\rm B}\over \mu_{\rm S}}-2\right) \simeq 2.5 ,\\
{\Delta \Omega \over   \Delta {\rm K} }
   &=& \left( {\mu_{\rm B}\over \mu_{\rm S}}-3\right)   \simeq 1.5  . 
\end{eqnarray}
These relation test the SHM and strangeness conservation, they 
do not differentiate model variants such as
 chemical equilibrium and chemical non-equilibrium.

\section{Final remarks}
 The total 
multiplicity yield, as well as the  yield of charmed particles, 
is originating predominantly in the 
early stage, primary parton reactions. For this
reason, we did not address absolute yields of hadrons, and,
similarly, 
cannot study the total charm yield in the context considered here.
However, one
may wonder, if   the appearance of small 
but distinguished charmed meson and baryon yield 
does not offer an interesting and independent 
probe of the properties of the 
hadronization state, or even it could influence the 
results presented.

A earlier study of charmed hadron production,  in chemical equilibrium
at $T=170$ MeV, has yielded interesting insights into the relative production
strengths of charmed mesons and baryons emerging at this 
particular hadronization temperature~\cite{Andronic:2003zv}. 
We will not enter into further discussion
of this subject, but note that:\\
a) for $\gamma_s > 2, $ these results imply   that $D_s(c\bar s)^+$ and
its antiparticle should be the dominant charmed hadron fraction;\\
b) the differences in yields between particles and antiparticles, 
seen in table 3 of Ref.\,\cite{Andronic:2003zv}, indicate that 
charm particle contribution to the asymmetries we studied are
totally negligible. It appears that measurement of relative yield 
of charmed mesons and baryons will reveal the charm hadronization 
condition, and we hope to return to this subject soon. 

The small relative 
number of charmed quarks and their even smaller
particle--anti-particle asymmetry assures that  these particles
do not impact any of the results we obtained. On the other 
hand, these particles offer another opportunity to explore
hadronization conditions. The formation of 
charmed hadrons is expected to occur prior to general hadronization,
considering the greater binding of charmed particles~\cite{Thews:2004sh}.

To summarize, 
we have presented a detailed study of the soft hadron 
production pattern at LHC.  We discussed, in turn, relative yields 
  such as $\phi/h$  and  $K/h$ which allow insights into the hadronization
conditions and help address questions related  of chemical equilibrium and
 non-equilibrium, as well as temperature of hadron freeze-out.

Perhaps the most interesting result to pursue experimentally is the 
large value of $\gamma_s$ expected in hadronization of over-saturated
QGP phase.
 The ratios, shown in  Figs.\,\ref{partrat} and \ref{baryrat}, are not 
very sensitive to the choices we made that determine
chemical potentials $\mu_{\rm B}$ and $\mu_{\rm S} $, 
they probe primarily the interplay
between the $\gamma_s$ and $T$. We note further that
most  these ratios, at the favored value  $\gamma_s \simeq 5$,
differ considerably  from the 
chemical equilibrium model expectations. The relatively large 
value of $\gamma_s$  we expect at LHC,  twice as large 
as our analysis finds at RHIC,  derives from a larger absolute density
of strangeness at hadronization of the deconfined phase, combined with 
lower prevailing temperature expected in deeper expansion supercooling. Indeed,
while at RHIC $\gamma_s^{\rm QGP}|_{\rm hadronization}\le   1$, 
at LHC we expect 
$\gamma_s^{\rm QGP}|_{\rm hadronization}\to  1.5$--$2$. The difference in the 
phase space size of QGP with HG than leads to $\gamma_s\simeq 5$.

We have further
presented an in depth discussion of particle--antiparticle asymmetries 
which address the challenge of chemical and strange quark chemical
potential measurement. Since the particle--antiparticle 
yield difference is small
compared to each individual yield, a
special effort will need to be made to acquire these difference ratios
$\Delta N_i\simeq $ 0.1--4\%  at the level of 10\% or
better. 

\vspace*{.5cm}
Work supported by a grant from the U.S. Department of
Energy  DE-FG02-04ER41318. 
LPTHE, Univ.\,Paris 6 et 7 is: Unit\'e mixte de Recherche du CNRS, UMR7589.
 

\vskip 0.3cm


\begin{thebibliography}{19}
\providecommand{\bibinfo}[2]{#2}

\bibitem{Huang:2005nd}
H.~Z.~Huang and J.~Rafelski,
AIP Conf.\ Proc.\  {\bf 756}, 210 (2005)
[arXiv:hep-ph/0501187].

\bibitem{Fromerth:2002wb}
  M.~J.~Fromerth and J.~Rafelski,
  ``Hadronization of the quark universe'',
  arXiv:astro-ph/0211346.

\bibitem{JJBook}
J.~Letessier and J.~Rafelski,
Cambridge Monogr.\ Part.\ Phys.\ Nucl.\ Phys.\ Cosmol.\  {\bf 18}, 1 (2002).

\bibitem{Letessier:2005qe}
  J.~Letessier and J.~Rafelski,
  ``Hadron production and phase changes in relativistic heavy ion collisions'',
  arXiv:nucl-th/0504028.

\bibitem{BDMRHIC}
P.~Braun-Munzinger, K.~Redlich and J.~Stachel,
``Particle production in heavy ion collisions'',
[arXiv:nucl-th/0304013], and references therein.
 
\bibitem{share} 
G.~Torrieri, W.~Broniowski, W.~Florkowski, J.~Letessier and J.~Rafelski,
[arXiv:nucl-th/0404083], Comp. Phys. Com. {\bf 167}, 229 (2005), see:\\
www.physics.arizona.edu/$\tilde{\phantom{.}}$torrieri/SHARE/share.html


\bibitem{Wheaton:2004qb}
  S.~Wheaton and J.~Cleymans,
  ``THERMUS: A thermal model package for ROOT'',
  arXiv:hep-ph/0407174; and 
  J.\ Phys.\ G {\bf 31}, S1069 (2005).

\bibitem{Petreczky:2001yp}
P.~Petreczky, F.~Karsch, E.~Laermann, S.~Stickan and I.~Wetzorke,
Nucl.\ Phys.\ Proc.\ Suppl.\  {\bf 106}, 513 (2002);\\
F.~Karsch and E.~Laermann,
``Thermodynamics and in-medium hadron properties from lattice QCD,''
arXiv:hep-lat/0305025, in: R.C. Hwa, R.C. (ed.) et al.: Quark gluon plasma III, pp. 
1-59 (Singapore 2004).


\bibitem{Uvarov:2001wv}
  V.~Uvarov,
  Phys.\ Lett.\ B {\bf 511}, 136 (2001)
  [arXiv:hep-ph/0105185].

\bibitem{Andronic:2003zv}
  A.~Andronic, P.~Braun-Munzinger, K.~Redlich and J.~Stachel,
  Phys.\ Lett.\ B {\bf 571}, 36 (2003),
  [arXiv:nucl-th/0303036].

\bibitem{destil}
  C.~Greiner, P.~Koch and H.~Stocker,
  Phys.\ Rev.\ Lett.\  {\bf 58} (1987) 1825;\\
  J.~Rafelski,
  Phys.\ Lett.\ B {\bf 190}, 167 (1987).

\bibitem{ChemCons}
J.~Letessier, A.~Tounsi, U.~W.~Heinz, J.~Sollfrank and J.~Rafelski,
Phys.\ Rev.\ D {\bf 51}, 3408 (1995),
[arXiv:hep-ph/9212210].

\bibitem{entro}
J.~Letessier, A.~Tounsi, U.~W.~Heinz, J.~Sollfrank and J.~Rafelski,
Phys.\ Rev.\ Lett.\  {\bf 70}, 3530 (1993).


\bibitem{bdepend} 
J.~Rafelski, J.~Letessier and G.~Torrieri,
``Centrality dependence of bulk fireball properties at RHIC'',
arXiv:nucl-th/0412072.
 

\bibitem{Thews:2004sh}
  R.~L.~Thews,
  J.\ Phys.\ G {\bf 31}, S641 (2005)
  [arXiv:hep-ph/0412323].

 

\end{thebibliography}
\end{document}